\documentclass[preprint,english,aps,prb,floatfix,preprintnumbers,showpacs,amsfonts,amssymb,superscriptaddress]{revtex4}
\usepackage{ae}
\usepackage{aecompl}
\usepackage[T1]{fontenc}
\usepackage[latin1]{inputenc}
\usepackage{amsmath}
\usepackage{graphicx}
\usepackage{amssymb}

\makeatletter
\usepackage{amscd}
\usepackage{bm}

\usepackage[portuguese]{babel}
\makeatother
\begin{document}

\title{The one-dimensional Kondo lattice model at quarter-filling}

\author{J.~C.~Xavier}

\affiliation{Instituto de F\'{\i}sica, Universidade Federal de Uberl\^{a}ndia,
Caixa Postal 593, 38400-902 Uberl\^{a}ndia, MG, Brazil}

\author{E.~Miranda}

\affiliation{Instituto de F\'{\i}sica Gleb Wataghin, University of Campinas -
UNICAMP, P. O. Box 6165, 13083-970 Campinas, SP, Brazil}

\date{\today}

\begin{abstract}
We revisit the problem of the quarter-filled one-dimensional Kondo
lattice model, for which the existence of a dimerized phase and a
non-zero charge gap had been reported in Phys. Rev. Lett. \textbf{90},
247204 (2003). Recently, some objections were raised claiming that
the system is neither dimerized nor has a charge gap. In the interest
of clarifying this important issue, we show that these objections
are based on results obtained under conditions in which the dimer
order is artificially suppressed. We use the incontrovertible dimerized
phase of the Majumdar-Ghosh point of the $J_{1}-J_{2}$ Heisenberg
model as a paradigm with which to illustrate this artificial suppression.
Finally, by means of extremely accurate DMRG calculations, we show
that the charge gap is indeed non-zero in the dimerized phase.
\end{abstract}


\maketitle

\section{Introduction}

\label{sec:Introduction}

The Kondo lattice model plays an important role in the discussion
of the physical properties of heavy fermion materials.\cite{hewson}
Therefore, an accurate determination of its properties and phase diagram
is an important task. In this respect, even the one-dimensional case
is of interest. Even though the Kondo lattice chain (KLC) has been
extensively analyzed and much is known about its properties\cite{Tsunetsugu}
(see also Ref. \onlinecite{gulacsi1dkondo}), a few controversies
still remain.

A few years ago we and our collaborators presented mostly numerical
evidence that the KLC has an insulating dimerized phase at quarter
filling.\cite{dimer,mirandaxavier04} Our analysis was mainly based
on numerical Density Matrix Renormalization Group (DMRG) calculations
with open boundary conditions (OBC). More recently, Hotta and Shibata
(HS) raised objections to our conclusions and suggested that the KLC
at quarter filling is neither dimerized nor exhibits a charge gap.\cite{shibatahotta05,shibatadimer}
They ascribed the putative errors to a lack of numerical precision
and to a misinterpretation of the data. Their analysis is mainly based
on modifications of the boundary conditions as compared to ours, either
by using an odd number of lattice sites and/or by imposing a shift
on the conduction electron site energies at the borders or by working
with periodic (PBC) and anti-periodic (APBC) boundary conditions.
Given these uncertainties and in view of the importance of this question,
it is our hope to shed some light on the origin of these differences,
while at the same time highlighting the intrinsic difficulties in
a numerical determination of the presence of dimerization.

As we will show below, if we use an odd number of sites, a very careful
interpretation of the data is necessary. We believe the discrepancy
between our results and those from HS stems from this fact. In order
to show this, we analyze a model with an incontrovertible dimerized
phase, namely, the e $J_{1}-J_{2}$ Heisenberg model at the Majumdar-Ghosh
(MG) point and by re-analyzing the KLC at quarter filling in light
of the latter results. Finally, we will also show extremely accurate
DMRG data confirming the existence of a charge gap.

Before presenting our results, let us first discuss briefly the difficulties
of determining numerically the presence of long-range order. Strictly
speaking, a spontaneously broken symmetry can only be realized in
the thermodynamic limit. Only in this limit is the order parameter
non-zero. No continuous symmetry can be spontaneously broken at \emph{finite}
temperature in one and two dimensions with short-ranged interactions,
as shown by the Mermin-Wagner-Hohenberg theorem.\cite{Auerbach} At
zero temperature, true long-range order is still possible for continuous
symmetry in two dimensions, while in one dimension only quasi-long-range
order can appear, i. e., the two point correlation function of the
order parameter decays as a power law. However, a \emph{discrete}
symmetry can be broken at zero temperature even in one dimension.
Such is the case of a dimerized phase, which can be characterized
by a broken discrete lattice translational symmetry. Although the
ordered phase is realized only in the thermodynamic limit, its signature
can be observed in finite systems. One possibility is a finite-size
scaling analysis of the order parameter correlation function. For
example, the Fourier transform of the spin-spin correlation function
at the ordering wave vector $S(\vec{q}^{*})$ can be studied as a
function of system size in order to detect long-range magnetic order.
If $S(\vec{q}^{*})$ (appropriately normalized) tends to a constant
as the system size grows, this signals the presence of long-range
magnetic order. The analogue in the case of dimerization would be
the dimer-dimer correlation function. Another possibility is to measure
the order parameter directly after the application of a small (infinitesimal)
symmetry-breaking field. If the order parameter remains non-zero as
the system grows, this is a sign of true long range order.

Our previous calculations followed the second route above. Since we
were interested in a possible broken lattice translation symmetry,
we used the open borders as small symmetry-breaking fields. However,
it is essential in this case that the lattice structure itself does
not suppress the possible dimer order, e.g., by working with a lattice
with an odd number of unit cells, which is unable to accommodate the
two-site sub-structure of the dimer order. Other modifications of
the boundaries can also be detrimental to the observation of dimerization.
As we will show, this accounts for the discrepancies between ours
and the results of HS. Furthermore, even when HS followed the first
route, their results will be shown not to be incompatible with the
presence of dimerization. 

It should be stressed that the numerical determination of long-range
order is often quite difficult as one is never sure whether a larger
lattice size will eventually show that an apparent order is actually
destroyed at longer length scales. However, it is very important to
work in such conditions as to allow the investigated order to at least
be possible. As we argue below, in the case of a dimerized phase,
extra care must taken not to frustrate the order from the start by
working with an odd number of sites or by artificially altering the
boundary conditions.

Given these uncertainties, we chose to show results for a system in
which the dimer order is well established, namely, the $J_{1}-J_{2}$
Heisenberg model for $J_{2}\gtrsim0.24J_{1}$. By showing the pitfalls
of a numerical determination of dimerization in this system, we hope
to both shed light on the previous works on the Kondo chain and to
bring to a more general audience what should be avoided in the investigation
of broken lattice translational symmetry with numerical methods.

We investigated the models above with DMRG \cite{white,white2} under
open boundary conditions (OBC). For the $J_{1}-J_{2}$ Heisenberg
model we used typically $m=400$ states per block. This number of
states kept in the truncation process is enough to give very precise
results, the discarded weight being typically about $10^{-10}$. However,
for the KLC we used a much larger number of states in order to obtain
precise results (up to $m=3500$). We have done $\sim$10-26 sweeps
and the discarded weight was typically $10^{-6}-10^{-10}$ in the
final sweep. The dimension of the superblock in the last sweep can
reach up to $27\times10^{6}$. This large dimension is due to the
fact that the center blocks in our DMRG procedure are composed of
8 states. As a consequence, keeping $m=3500$ states per block in
the KLC is analogous to keeping $m'=4\times3500$ states in the $J_{1}-J_{2}$
Heisenberg model.

This paper is organized as follows. In Section~\ref{sec:The-Manjumdar-Gosh-point},
we will use the MG point of the $J_{1}-J_{2}$ Heisenberg model as
a paradigm of a dimerized phase and important insight will be gained
as to the effect of the use of an odd number of lattice sites. In
Section~\ref{sec:Dimerization-in-the}, we will analyze the KLC for
both odd and even number of sites and will show that indeed our numerical
results point to the existence of a true long-ranged dimerized phase
at quarter-filling. In Section~\ref{sec:The-charge-gap}, through
extremely accurate DMRG calculations we will show that the charge
gap is indeed larger than zero in the dimerized phase. Finally, we
will present our conclusions.

\section{The Majumdar-Gosh point}

\label{sec:The-Manjumdar-Gosh-point}

One of the best established examples of a dimerized phase occurs in
the Heisenberg model with nearest ($J_{1}$) and next-nearest neighbor
($J_{2}$) interactions\[
H=\sum_{j=1}^{L-1}\left(J_{1}\mathbf{s}_{j}\cdot\mathbf{s}_{j+1}+J_{2}\mathbf{s}_{j}\cdot\mathbf{s}_{j+2}\right),\]
where $\mathbf{s}_{j}$ is a spin-$1/2$ operator at site $j$. This
model is known to show dimerization for $J_{2}/J_{1}\equiv\alpha>\alpha_{c}\approx0.24$.\cite{haldanedimer,okamotonomura92,bursilletal,whiteaffleck96}
A particularly simple point of the dimerized phase is the MG point
$\alpha=0.5$. The ground state of the infinite system can be shown
to be composed of independent singlets formed out of neighboring pairs
of spins.\cite{majumdarghosh1,majumdarghosh2,majumdarghosh3} This
much simpler structure, as compared to $\alpha\neq0.5$, corresponds
to a correlation length of one lattice spacing for the connected dimer-dimer
correlation function. Evidently, there are two equivalent ways of
realizing the broken translational-symmetry ground state by choosing
the dimers to lie on pairs which are one lattice spacing apart. The
exact ground state energy follows trivially from this structure and
is given by $E_{0}/L=-3/8J_{1}$.

If the model is now analyzed on a finite lattice with an \emph{even}
number of sites and OBC, \emph{one} of the two equivalent ground states
can be easily accommodated, the other becoming now an excited state.
However, for an \emph{odd} number of sites and OBC, \emph{none} of
the above ground states can be realized, as the two-site sub-structure
cannot be accommodated in an odd-sized lattice. In other words, the
additional site frustrates the dimerization. This not specific to
the MG point, but applies to \emph{any} dimerized phase. The energies
at the MG point can no longer be obtained exactly when $L$ is odd,
although some variational estimates can be made.\cite{Shastrysoliton}
Below we show some energies obtained numerically for odd $L$. We
note that for PBC (for which there is not explicit broken translational
symmetry) \emph{}and finite even $L$, the ground state is, in general,
a superposition of the two equivalent broken-symmetry states and does
\emph{not} show spontaneously broken translational invariance. This
is to be expected, since symmetries can only be spontaneously broken
in the thermodynamic limit. 

Note that for OBC the average nearest-neighbor spin-spin correlation
$D\left(j\right)=<\mathbf{s}_{j}\cdot\mathbf{s}_{j+1}$>, is not a
constante (depends of $j$) for a \emph{finite system}, since translational
symmetry is \emph{explicitly} broken by the chain ends, which act
effectively like small symmetry-breaking fields as $L\to\infty$. 

Let us define the dimer order parameter as $|O_{j}|=|D(j)-D(j+1)|$.
For OBC, the signature of a true long-range dimer order should be
detected by plotting \emph{$|O_{L/2}|$} versus $1/L$. The dimer
order exists if $\lim_{{\normalcolor L\rightarrow\infty}}|O_{L/2}|>0$,
since the boundary-generated symmetry-breaking field becomes infinitesimal
and the thermodynamic limit is enforced.

\begin{figure}[t]
\begin{centering}\includegraphics[scale=0.26]{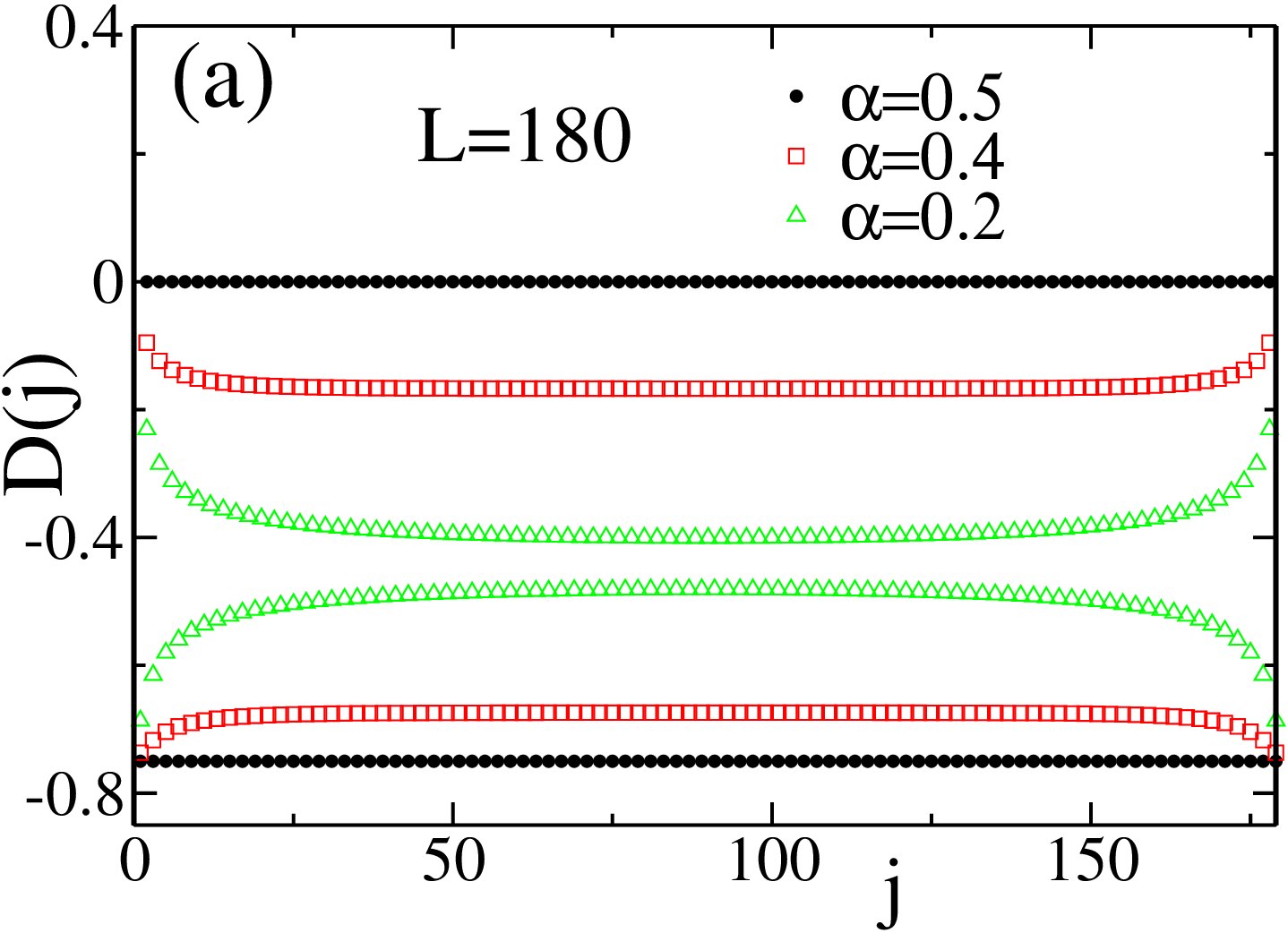}\par\end{centering}

\begin{centering}\includegraphics[scale=0.26]{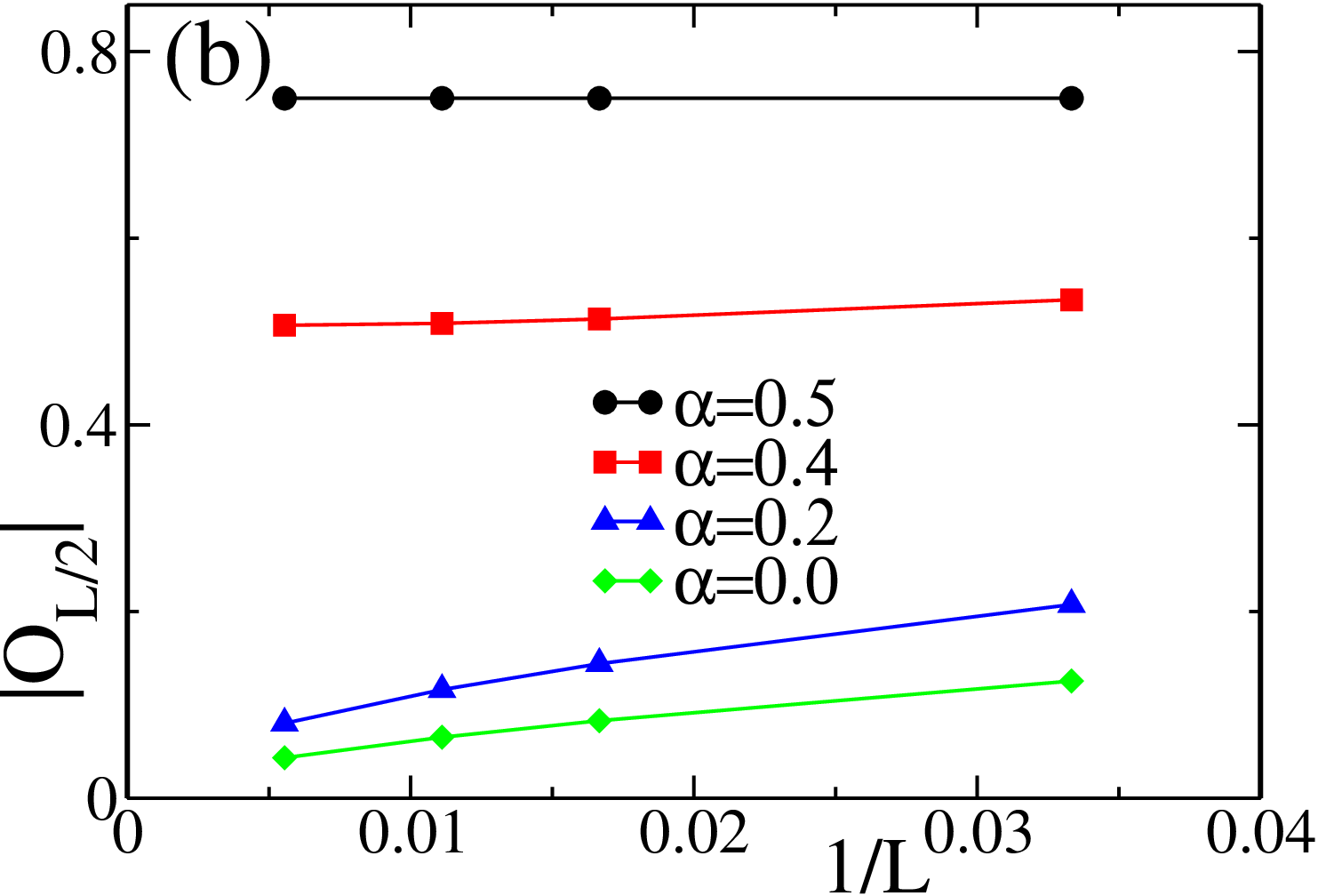}\par\end{centering}

\caption{\label{fig1} (Color online). (a) The nearest-neighbor spin-spin
correlation $D\left(j\right)$ versus \emph{}distance from the lattice
boundary for the $J_{1}-J_{2}$ Heisenberg model with $L=180$ and
some values of $\alpha$. (b) The dimer order parameter at the chain
center $|O_{L/2}|$ as a function of $1/L$ for lattices with \emph{even}
numbers of sites for the same model.}
\end{figure}

Let us first consider the $J_{1}-J_{2}$ Heisenberg model with an
\emph{even number of sites}. In Fig.~\ref{fig1}(a), $D(j)$ is shown
for the $J_{1}-J_{2}$ Heisenberg model with OBC for some values of
$\alpha$ and $L=180$ (note that there are $L-1$ data points for
$D\left(j\right)$ in an $L$-site lattice). As we mentioned before,
with OBC $D(j)$ oscillates. In particular, for the MG point ($\alpha=0.5$)
$D(j)$ is zero for even $j$ and $-3/4$ for odd $j$, as expected.
The fact that $D(j)$ exhibits a robust oscillation in the middle
of the system suggests that a dimer order may exist. In order to establish
the dimer phase, we now have to plot \emph{$|O_{L/2}|$} as a function
of $1/L$. This is shown in Fig.~\ref{fig1}(b), where we clearly
see that the order parameter tends to a non-zero value only for $\alpha>0.2$.
Though not intended to determine the critical value $\alpha_{c}\sim0.24$
, this plot shows that it is possible to numerically determine \emph{the
presence or the absence} of long-ranged dimer order using the second
route mentioned before.

\begin{figure}[t]
\begin{centering}\includegraphics[scale=0.35]{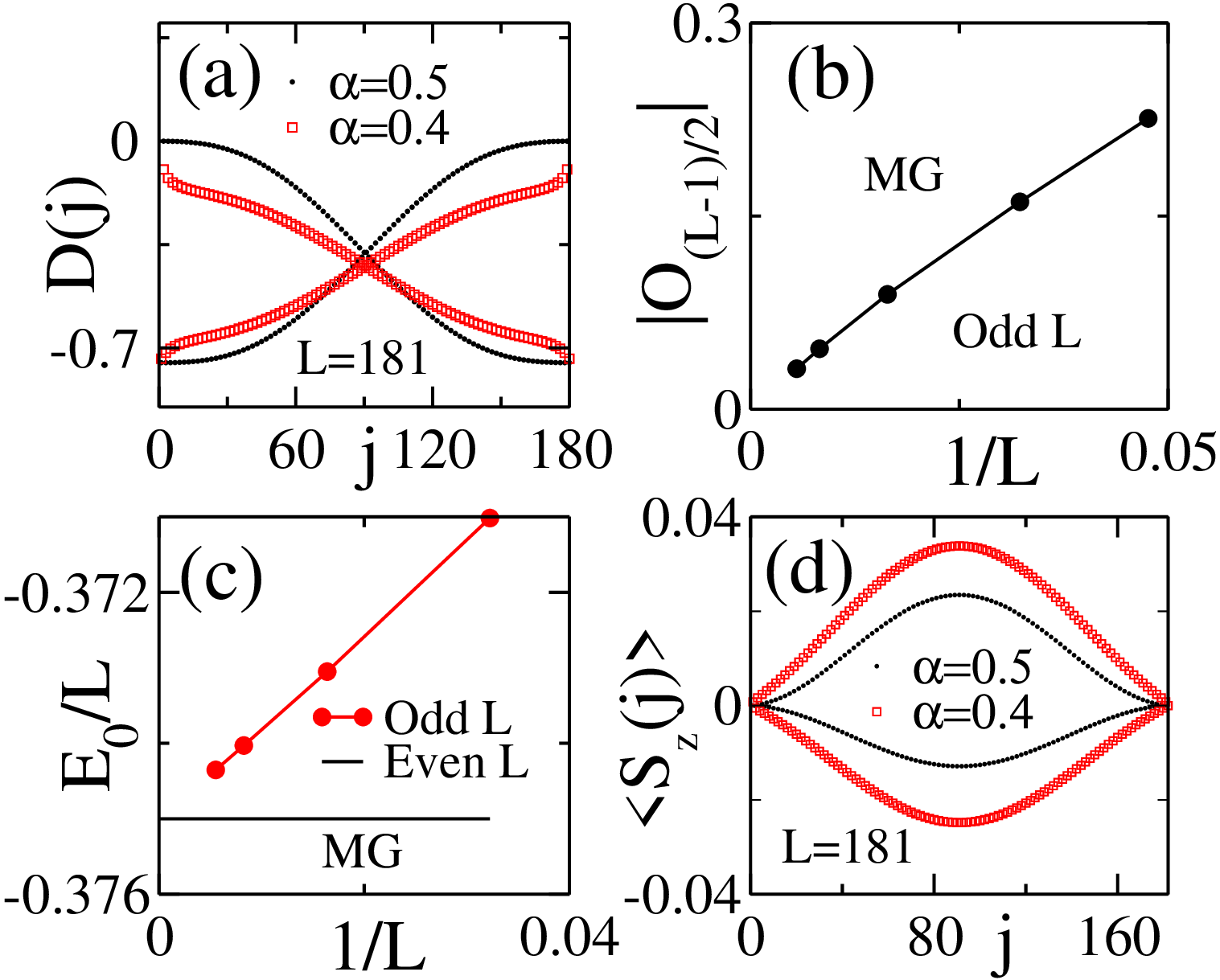}\par\end{centering}

\caption{\label{fig2} (Color online). Some expectation values and energies
for the $J_{1}-J_{2}$ Heisenberg model ($J_{2}/J_{1}\equiv\alpha$).
(a) $D(j)$ \emph{vs} distance from the lattice boundary for $L=181$
at the Majumdar-Ghosh point ($\alpha=0.5$) and at $\alpha=0.4$;
(b) $|O_{(L-1)/2}|$ as a function of $1/L$ for lattices with \emph{odd}
numbers of sites at the MG point; (c) the ground state energy per
site $E_{0}/L$ \emph{vs} $1/L$ for even and odd $L$ at the MG point.
(d) $\left\langle S_{z}(j)\right\rangle $ \emph{vs} lattice site
for $L=181$, showing the solitonic excitation, at the MG point and
at $\alpha=0.4$.}
\end{figure}

Now, let us investigate the same model with an \emph{odd number of
sites}. In Fig.~\ref{fig2}(a), we present the average nearest-neighbor
spin-spin correlation $D(j)$ at the MG point ($\alpha=0.5$) and
at $\alpha=0.4$ as a function of $j$ for \emph{odd} system sizes
with OBC. As expected, $D(j)$ oscillates. However, we clearly see
that this oscillation seems to vanish in the middle of the lattice.
At first sight, this seems to suggest that in the thermodynamic limit
$D(j)$ does not oscillate, indicating the absence of a dimerized
phase in both cases. In order to check this, we present in Fig.~\ref{fig2}(b)
$|O_{(L-1)/2}|$ \emph{}versus $1/L$ for \emph{odd} $L$ (only for
the MG point). As can be seen, $|O_{j}$| indeed tends to zero in
the thermodynamic limit, indicating, at first thought, the absence
of a dimerized phase. This result is in contradiction with the well
known dimerized phase of this model both at the MG point and at $\alpha=0.4$.
How is it possible? The explanation lies in the frustration of translational
symmetry breaking induced by the last site of an odd-sized lattice,
as explained above. In fact, the ground state at the MG point (and
at $\alpha=0.4$) with an odd number of sites has a solitonic spin-$1/2$
excitation, which is delocalized and acts to suppress the dimer order.\cite{afflecksoliton,Shastrysoliton}
An estimate of the ground-state energy at the MG point can be obtained
assuming that the wave function for odd $L$ is given by a free solitonic
excitation.\cite{Shastrysoliton} In fact, numerical data support
this picture.\cite{afflecksoliton}

In Fig.~\ref{fig2}(c), the ground state energy per site, for odd
and even $L$, is presented as function of $1/L$ (only at the MG
point). Clearly, the energy per site for odd $L$ is larger than for
even $L$. This is due to the fact that there is a gap for the creation
of the solitonic excitation. In Fig.~\ref{fig2}(d), we show $<S_{z}(i)>$
at the MG point and at $\alpha=0.4$, for a system size of $L=181$
(similar results were obtained in reference {[}\onlinecite{afflecksoliton}]
for $L=101$). The delocalization of the soliton is clear in both
cases and its envelope can be modeled as a free quantum particle in
a box of size $L$.\cite{afflecksoliton} The soliton has a finite
extent for $\alpha\neq0.5$, due to the larger dimer correlation length,
and this picture becomes less accurate.\cite{afflecksoliton} As we
argue in the next Section, the frustration of the dimer order in an
odd-sized lattice is the origin of the discrepancy between ours and
HS´s work.

\section{Dimerization in the quarter-filled Kondo Lattice Chain}

\label{sec:Dimerization-in-the}

Having obtained some insight into the ground state wave function at
the MG point with an odd number of sites, we consider now the KLC
with OBC given by 

\[
H=-\sum_{i=1,\sigma}^{L-1}(c_{i,\sigma}^{\dagger}c_{i+1\sigma}^{\phantom{\dagger}}+\mathrm{H.}\,\mathrm{c.})+J\sum_{j=1}^{L}\mathbf{S}_{j}\cdot\mathbf{s}_{j},\]
where $c_{j\sigma}$ annihilates a conduction electron in site $j$
with spin projection $\sigma$, $\mathbf{S}_{j}$ is a localized spin-$1/2$
operator, $\mathbf{s}_{j}=\frac{1}{2}\sum_{\alpha\beta}c_{j,\alpha}^{\dagger}\bm{\sigma}_{\alpha\beta}c_{j,\beta}^{\phantom{\dagger}}$
is the conduction electron spin density operator and $\bm{\sigma}_{\alpha\beta}$
are Pauli matrices. Here, $J>0$ is the Kondo coupling constant between
the conduction electrons and the local moments and the hopping amplitude
was set to unity to fix the energy scale.

\begin{figure}
\begin{centering}\includegraphics[scale=0.35]{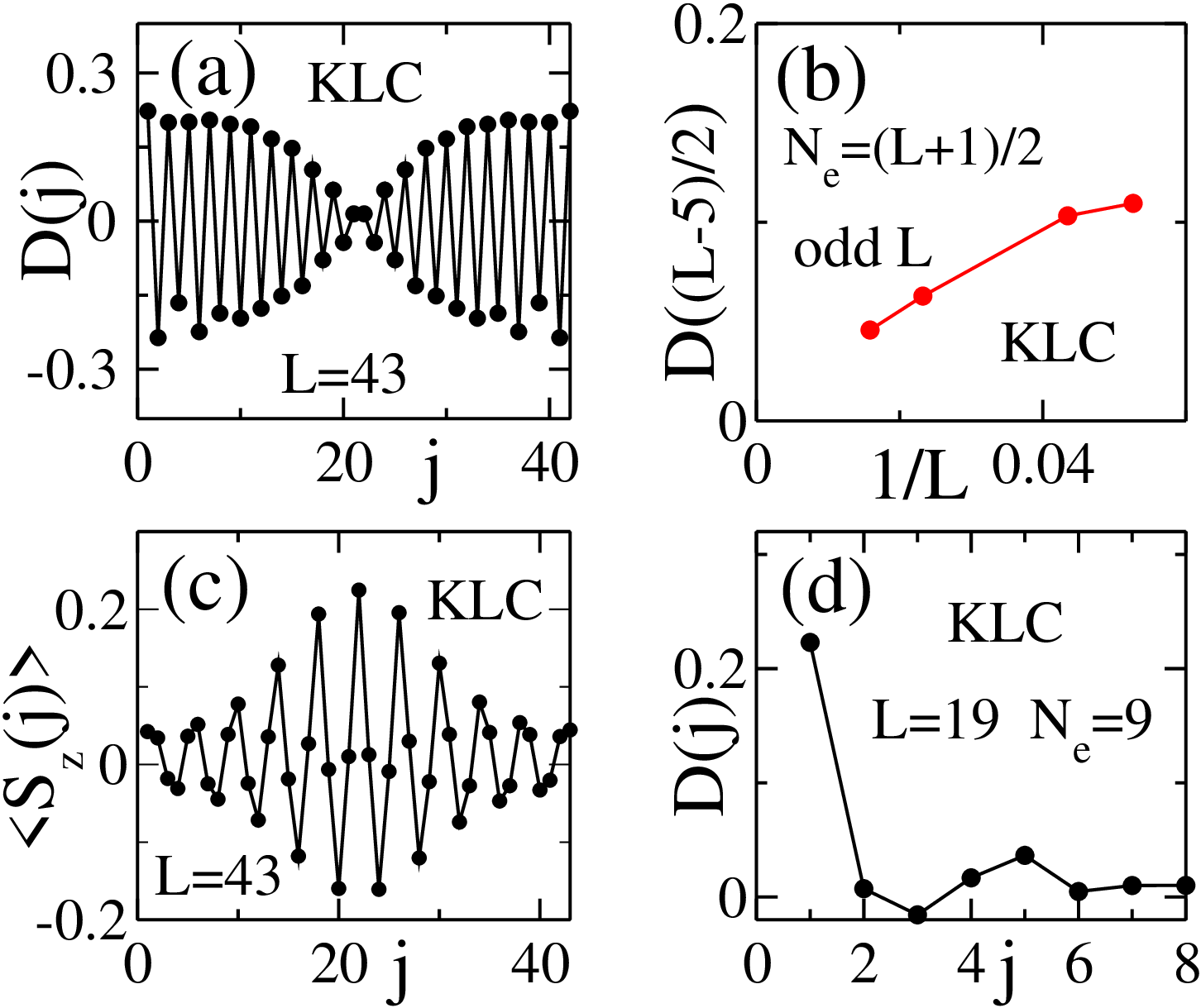}\par\end{centering}

\caption{\label{fig3} (Color online) (a) The dimer order parameter $D(j)$
\emph{vs} distance to the boundary for the KLC, $J=0.5$, $N_{e}=22$,
and $L=43$ (only half the sites are shown). (b) The dimer order parameter
at the chain center ($D((L-5)/2)$) as a function \emph{}$1/L$ for
odd $L$ and the same parameters as in (a); (c)$<S_{z}(j)>$ \emph{vs}
$j$ for the KLC and the same parameters as in (a); (d) Same as (a)
but for $L=19$ and $N_{e}=9$. }
\end{figure}

In our previous work,\cite{dimer} we showed that the KLC at quarter-filling
is dimerized from $\lim_{{\normalcolor L\rightarrow\infty}}|D(L/2)|>0$
for even $L$. Note that for the KLC, $D(j)$ oscillates in sign and,
for this reason, we can use $D(j)$ as the dimer order parameter.
In analogy to the MG point, we can destroy the dimerization by breaking
the 2-site sub-structure of the dimers. For example, we can add/remove
just one site. In this case, keeping the electronic density as close
as possible to quarter-filling, the electronic densities are $n=\frac{1}{2}\pm\frac{1}{2L}$.
For odd $L$ the ground-state total spin is not zero. If the number
of conduction electrons $N_{e}$ is even, it is 1/2. On the other
hand, for odd $N_{e}$ the total spin of the ground state is 1(2)
if $L=2N_{e}-1$ ($L=2N_{e}+1$). It is interesting that if we keep
the lattice size even and add/remove just one electron in the conduction
band, we may also destroy the dimerization, as discussed below.

\begin{figure}
\begin{centering}\includegraphics[scale=0.26]{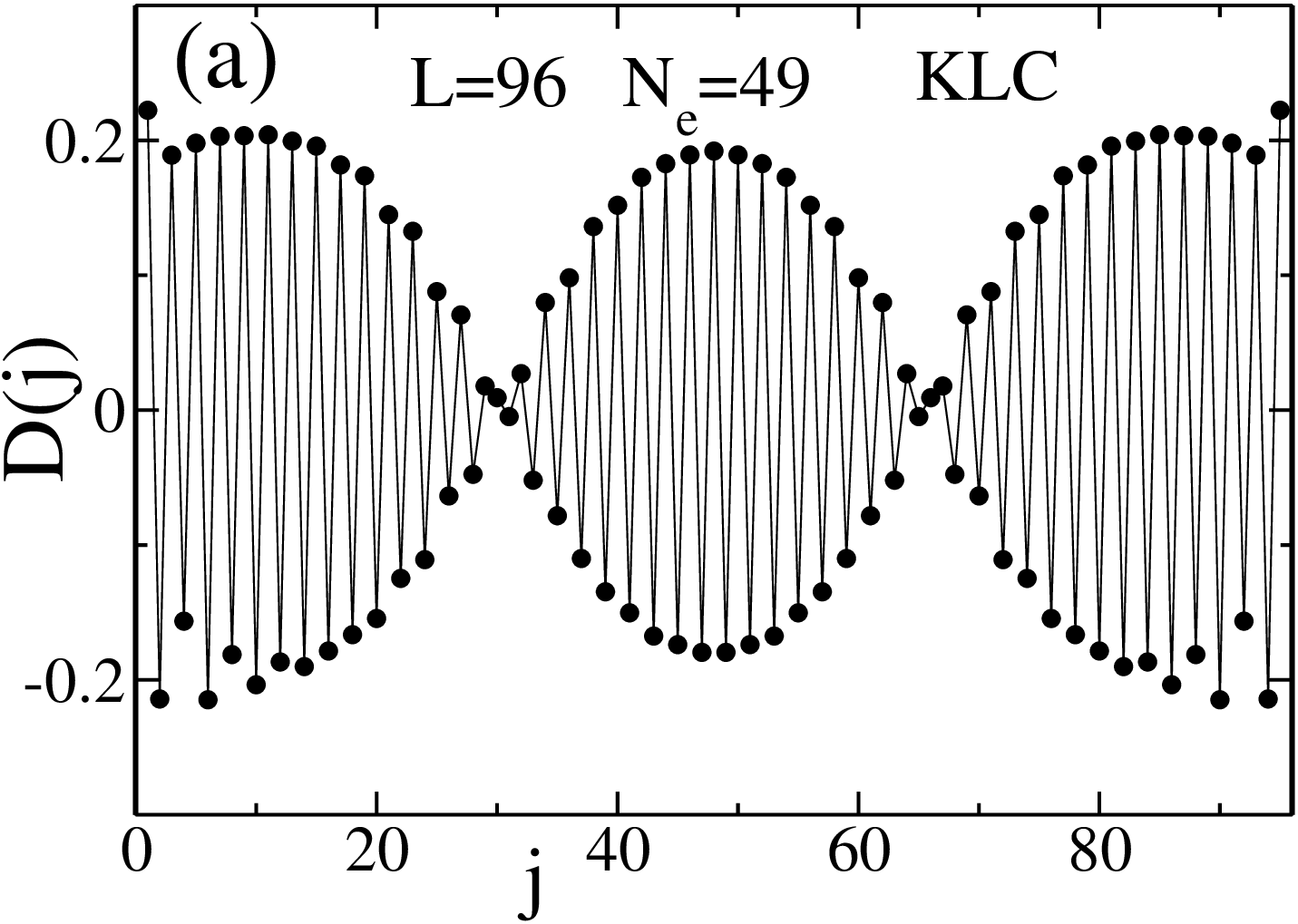}\par\end{centering}

\begin{centering}\includegraphics[scale=0.26]{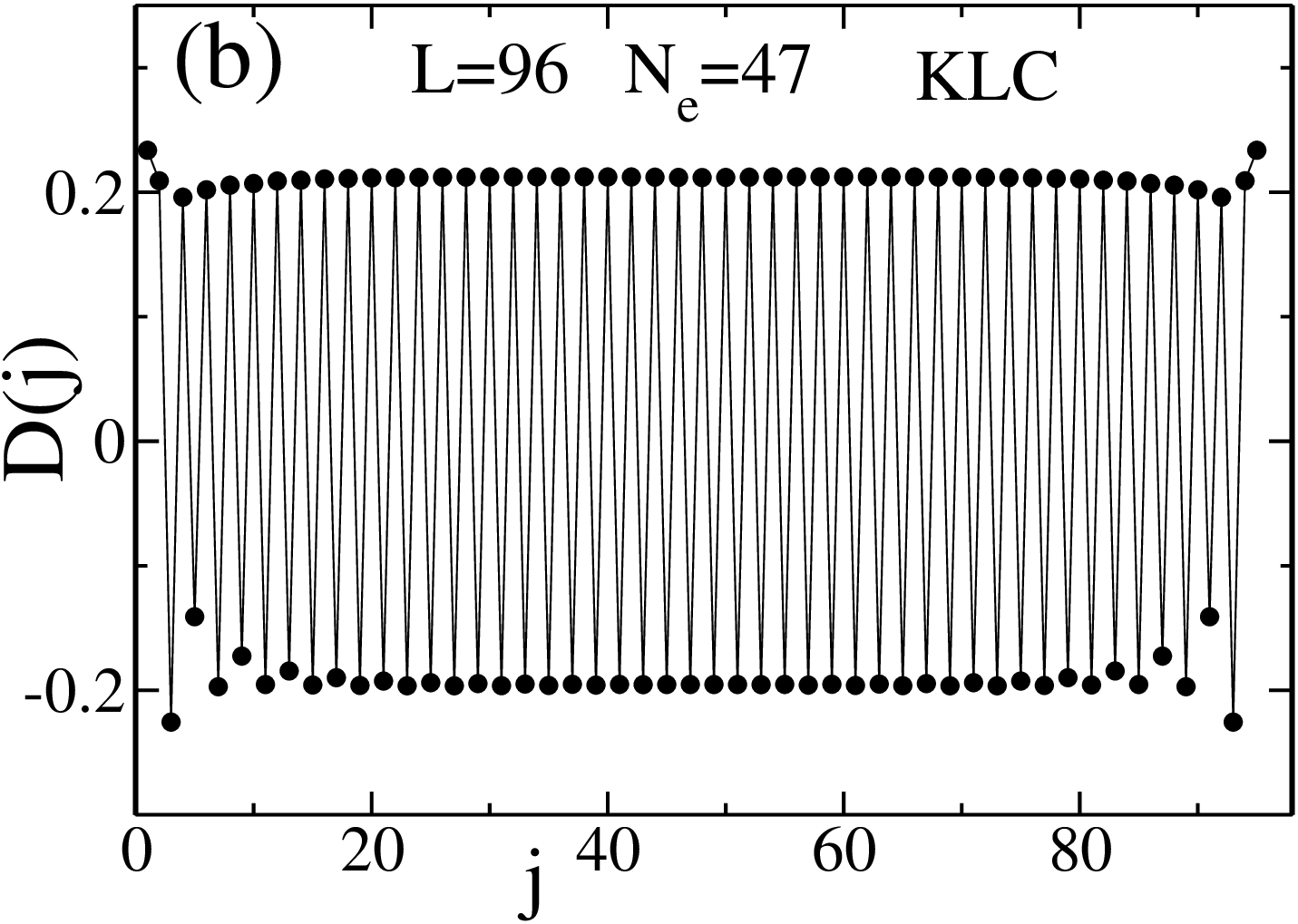}\par\end{centering}

\caption{\label{fig4} The dimer order parameter $D(j)$ \emph{vs} distance
to the boundary for the KLC and $J=0.5$: (a) $L=96$ and $N_{e}=49$
(b) $L=96$ and $N_{e}=47$. }
\end{figure}

In Fig.~\ref{fig3}(a), the order parameter $D(j)$ is shown as a
function of the distance to the boundary for the KLC with $J=0.5$
and $N_{e}=\frac{L+1}{2}$. As in the case of the MG point, the order
parameter seems to decrease away from the boundary. This result suggests
that the dimerization does not exist in the thermodynamic limit for
the KLC with odd $L$. Indeed, as shown in Fig.~\ref{fig3}(b), the
dimer order parameter measured at the center of the lattice tends
to zero as we increase the system size. We also show in Fig.~\ref{fig3}(c)
$<S_{z}(j)>$ as a function of the lattice site. Note the similarity
with the MG point (Fig.~\ref{fig2}(d)). These results suggest that
the ground state of KLC with odd $L$ also possesses a solitonic excitation,
as at the MG point with odd $L$. If $L$ is odd and $N_{e}=\frac{L-1}{2}$
the dimer order parameter decays much faster away from the edges,
as can be seen in Fig.~\ref{fig3}(d) for a representative set of
data. Although the energy converges quite rapidly in this case, the
convergence of the correlations is slower, and for this reason we
restrict our results to {}``small'' lattice sizes. 

For even $L$ and odd $N_{e}$ the total spin of the ground state
is $3/2$. For even $L$ and $N_{e}=\frac{L}{2}+1$, the dimerization
also seems to decay away from the boundary, although it does so in
an oscillatory fashion, as seen in Fig.~\ref{fig4}(a). One would
perhaps naively think that also for $N_{e}=\frac{L}{2}-1$ and even
$L$ the dimerization might not exist. However, as shown in Fig.~\ref{fig4}(b),
the order is robust in this case. It is likely that the solitons are
localized at the boundaries in the case of $N_{e}=\frac{L}{2}-1$.

\section{The charge gap in the Kondo lattice chain}

\label{sec:The-charge-gap}

Finally, we investigate the charge gap of the KLC at quarter-filling.
We define the charge gap as $\Delta=E(N+2)+E(N-2)-2E(N)$, where $E(M)$
is the ground state energy for $M$ conduction electrons. In our previous
work \cite{dimer} we presented numerical results with DMRG keeping
$m=800$ indicating that there exists a finite charge gap at quarter
filling in the thermodynamic limit. We also used bosonization arguments
to show that, if the KLC has a dimerized ground state at quarter-filling,
then a charge gap could exist. However, Hotta and Shibata raised objections
to our previous results arguing for a lack of precision in our numerical
data for the coupling value $J=0.5$. It should be stressed that these
authors only showed results for $J=1$ (and $m=1100$). Here, we present
new DMRG results, keeping up to $m=3500$ states, showing that their
objections are unfounded.

\begin{figure}[t]
\begin{centering}\includegraphics[scale=0.26]{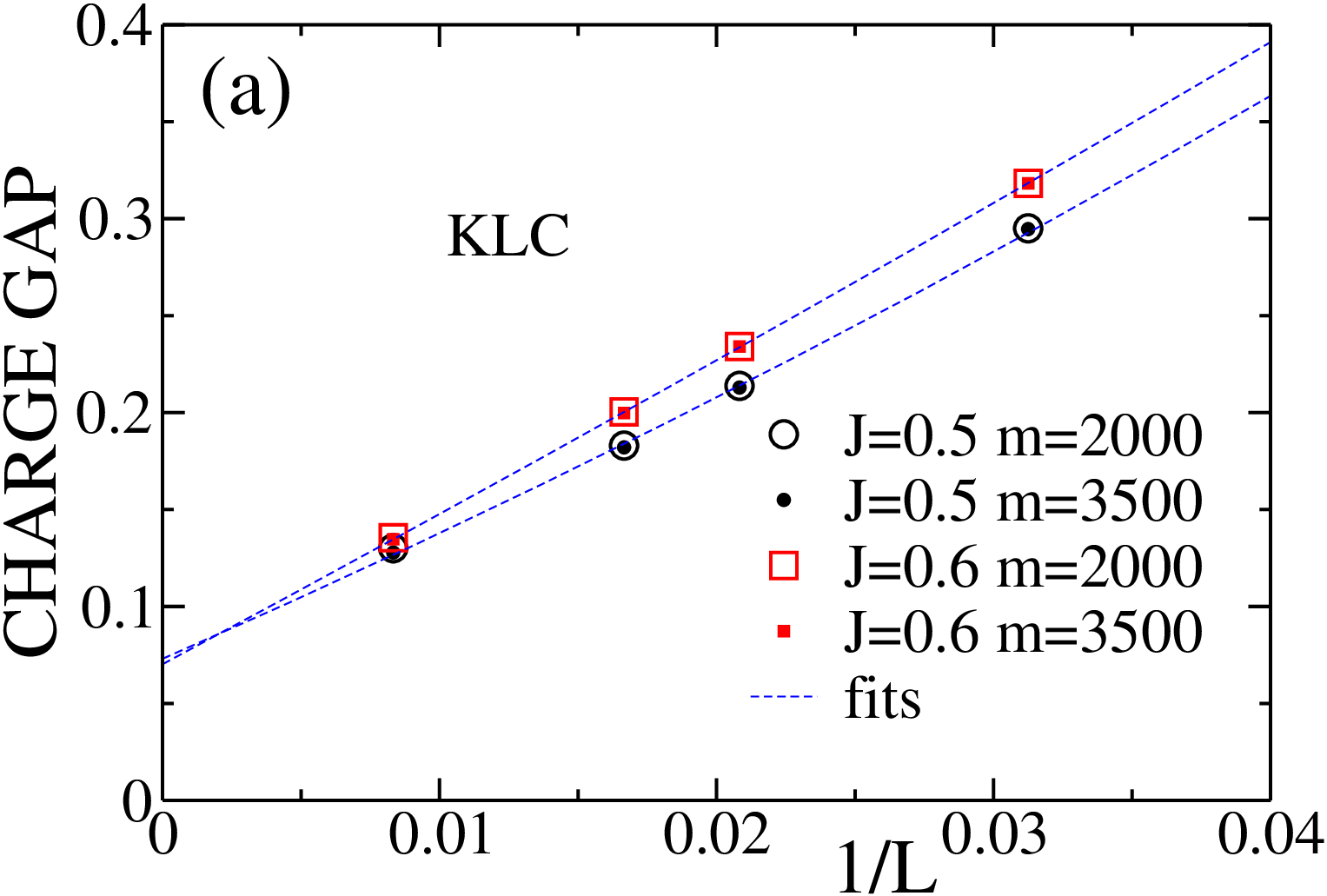}\par\end{centering}

\begin{centering}\includegraphics[scale=0.26]{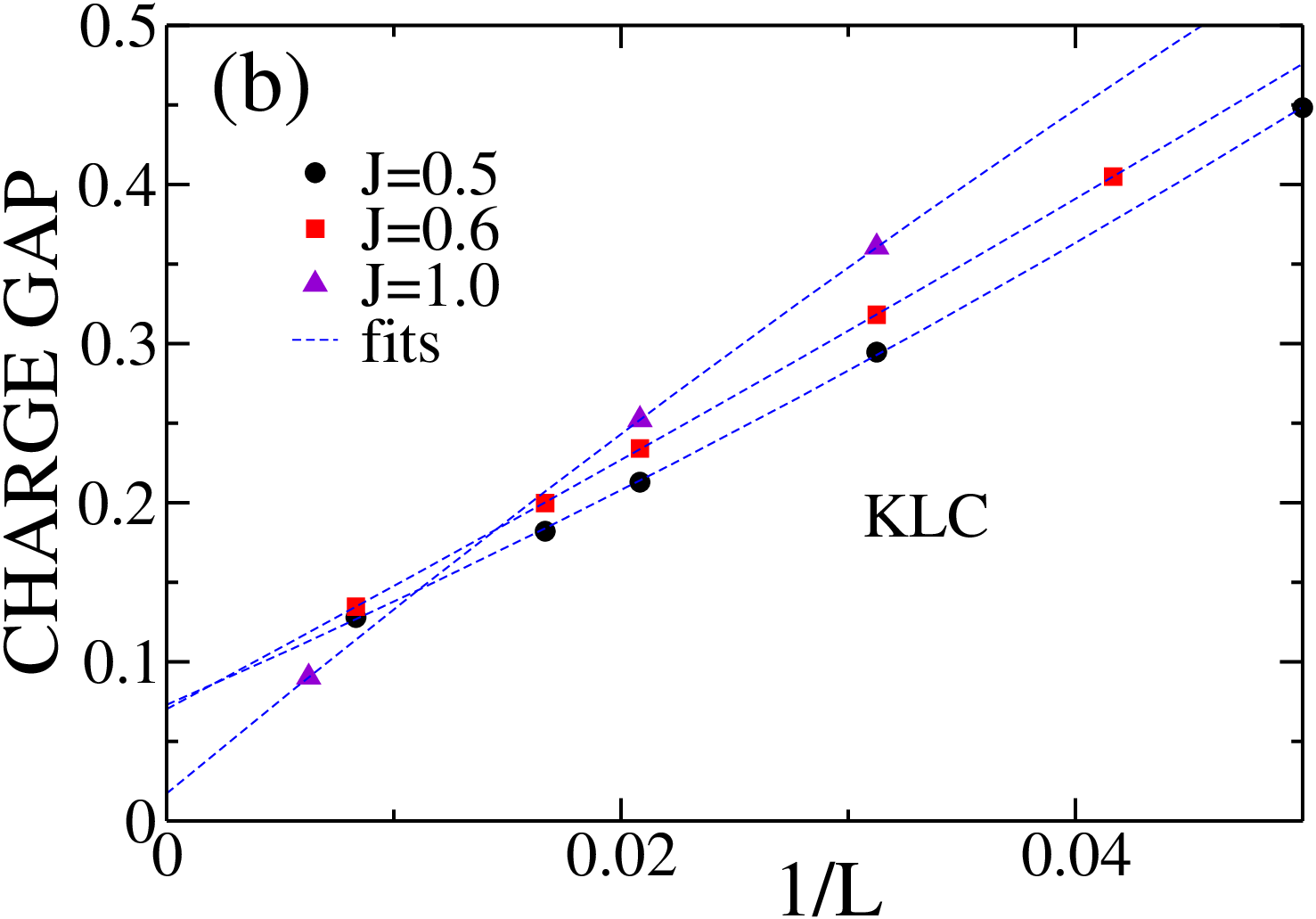}\par\end{centering}

\caption{\label{fig5} (Color online). Charge gaps \emph{$\Delta$} for the
KLC at quarter filling. (a) $\Delta$ \emph{vs} $1/L$ for $J=0.5$
and $J=0.6$ and two values of the DMRG truncation $m$. The dotted
lines are fits (see text for details). (b) $\Delta$ \emph{vs} $1/L$
for $J=0.5$, $0.6$ and $1$. }
\end{figure}

In Fig.~\ref{fig5}(a), the charge gaps for coupling constants $J=0.6$
and $J=0.5$ are shown as functions of $1/L$ for two different values
of $m$. As can be observed, there is very little difference between
the charge gaps $\Delta_{L}(m=3500)$ and $\Delta_{L}(m=2000)$. We
also show a fit of our data with $m=3500$ with the function $\Delta_{L}=\Delta_{\infty}+a_{0}/L+a_{1}/L^{2}$.
The charge gaps clearly tend to a non-zero value. Our fit gives $\Delta_{\infty}=0.070$
and $\Delta_{\infty}=0.072$ for $J=0.6$ and $J=0.5$, respectively.
However, HS did not present results for $J=0.5$ or $J=0.6$ where
the charge gaps are expected to be larger than for $J=1$, since the
RKKY interaction which we argued induces the dimerization\cite{dimer}
is expected to dominate the {}``Kondo effect'' for small $J$. Indeed,
this is what we obtain, see Fig.~\ref{fig5}(b). The results of Hotta
and Shibata are confined to $J=1$, where the charge gap is quite
small as we see in Fig.~\ref{fig5}(b). Finally, the extrapolations
for densities away from half-filling presented in our previous papers
give $\Delta_{\infty}<0.002$, which are much smaller than the charge
gaps found for quarter filling and beyond our precision.

\section{Conclusions}

\label{sec:Conclusions}

In the above, we have shown how an uncritical finite-size analysis
can lead to wrong conclusions in the incontrovertibly dimerized phase
of the MG point. A very similar situation was found in the case of
the quarter-filled KLC, where very different results are obtained
for even and odd numbers of lattice sites. This shows how care must
be taken in a numerical determination of dimerization, specially with
the number of sites and the boundary conditions.

In their work, HS used different boundary conditions (OBC and APBC),
odd and even numbers of sites and even modified the site energies
at the edges of the chain.\cite{shibatahotta05,shibatadimer} As argued
above, any of these \emph{ad hoc} changes can severely affect the
dimer order and in many cases suppress it to zero. The use of OBC
with an even number of sites, thus commensurate with a possible dimerized
sub-structure, is an unbiased way of probing the long-range nature
of the broken lattice translational symmetry in the thermodynamic
limit. 

Another unbiased strategy is the investigation of the long-distance
behavior of the dimer-dimer correlation function $C\left(i-j\right)\equiv\left\langle \left(\mathbf{S}_{i}\cdot\mathbf{S}_{i+1}\right)\left(\mathbf{S}_{j}\cdot\mathbf{S}_{j+1}\right)\right\rangle $
with PBC and even $L$. This procedure is much less convenient in
a DMRG calculation, since the DMRG is much less efficient under PBC
or APBC and one is thus confined to small system sizes. HS showed
also results under APBC. Although we are not certain whether the additional
phase introduced under APBC suppresses dimerization or not, it is
clear that their results is fully compatible with a dimerized ground
state. Indeed, for $J=1$ and $L=120$ our calculations give a dimer
order parameter $\left|\left\langle d\left(j\right)\right\rangle \right|=\left|\left\langle \mathbf{S}_{j}\cdot\mathbf{S}_{j+1}\right\rangle \right|\approx0.128$.
This corresponds to $\left|\left\langle o_{j}\right\rangle \right|\equiv\left|<d\left(j\right)-d\left(j+1\right)>\right|\approx0.256$
and to $\lim_{\left|i-j\right|\to\infty}\left|\left\langle o_{i}o_{j}\right\rangle \right|=\left|\left\langle o_{i}\right\rangle \right|^{2}\approx0.066$.
It can be seen from the data reported by HS\cite{shibatahotta05,shibatadimer}
that the asymptotic behavior of their dimer-dimer correlation function
seems to be just converging to about this value for $L=32$ and APBC.
Thus, although it is hard to say whether they have reached the asymptotic
bulk value, their results for APBC are certainly not incompatible
with true long-ranged dimer order. It would be interesting to investigate
this question under PBC and larger system sizes, in order to have
a complete picture. Unfortunately, with the DMRG this may well turn
out to be impossible with currently available computer power.

Finally, we have also shown that even the most accurate DMRG calculations
($m=3500$) point to the existence of a non-zero charge gap at quarter-filling.
HS confined their results to the $J=1$ case (and $m=1100$), where
this gap is very tiny (also from the $m=3500$ results), which makes
the analysis more difficult. By working at smaller coupling constant
values, we were able to accurately extrapolate to the thermodynamic
limit and found that the charge gap indeed tends to a non-zero value.

\begin{acknowledgments}
This work was supported by Brazilian agencies FAPEMIG (JCX) through
grant CEX-447/06 and CEX-3633/07, and CNPq through grants 305227/2007-6
(EM), 472635/2006-9 (JCX), and 303072/2005-9 (JCX).
\end{acknowledgments}

\begin{thebibliography}{18}
\expandafter\ifx\csname natexlab\endcsname\relax\def\natexlab#1{#1}\fi
\expandafter\ifx\csname bibnamefont\endcsname\relax
  \def\bibnamefont#1{#1}\fi
\expandafter\ifx\csname bibfnamefont\endcsname\relax
  \def\bibfnamefont#1{#1}\fi
\expandafter\ifx\csname citenamefont\endcsname\relax
  \def\citenamefont#1{#1}\fi
\expandafter\ifx\csname url\endcsname\relax
  \def\url#1{\texttt{#1}}\fi
\expandafter\ifx\csname urlprefix\endcsname\relax\def\urlprefix{URL }\fi
\providecommand{\bibinfo}[2]{#2}
\providecommand{\eprint}[2][]{\url{#2}}

\bibitem[{\citenamefont{{A. C. Hewson}}(1993)}]{hewson}
\bibinfo{author}{\bibnamefont{{A. C. Hewson}}}, \emph{\bibinfo{title}{The
  {K}ondo {P}roblem to {H}eavy {F}ermions}} (\bibinfo{publisher}{Cambrige
  University Press, England}, \bibinfo{address}{Cambridge},
  \bibinfo{year}{1993}).

\bibitem[{\citenamefont{{H. Tsunetsugu} et~al.}(1997)\citenamefont{{H.
  Tsunetsugu}, {M. Sigrist}, and {K. Ueda}}}]{Tsunetsugu}
\bibinfo{author}{\bibnamefont{{H. Tsunetsugu}}},
  \bibinfo{author}{\bibnamefont{{M. Sigrist}}}, \bibnamefont{and}
  \bibinfo{author}{\bibnamefont{{K. Ueda}}}, \bibinfo{journal}{Rev. Mod. Phys.}
  \textbf{\bibinfo{volume}{69}}, \bibinfo{pages}{809} (\bibinfo{year}{1997}).


\bibitem[{\citenamefont{{M. Gulacsi}}(2004)}]{gulacsi1dkondo}
\bibinfo{author}{\bibnamefont{{M. Gulacsi}}}, \bibinfo{journal}{Adv. Phys.}
  \textbf{\bibinfo{volume}{53}}, \bibinfo{pages}{769} (\bibinfo{year}{2004}).


\bibitem[{\citenamefont{Xavier et~al.}(2003)\citenamefont{Xavier, Pereira,
  Miranda, and Affleck}}]{dimer}
\bibinfo{author}{\bibfnamefont{J.~C.} \bibnamefont{Xavier}},
  \bibinfo{author}{\bibfnamefont{R.~G.} \bibnamefont{Pereira}},
  \bibinfo{author}{\bibfnamefont{E.}~\bibnamefont{Miranda}}, \bibnamefont{and}
  \bibinfo{author}{\bibfnamefont{I.}~\bibnamefont{Affleck}},
  \bibinfo{journal}{Phys. Rev. Lett.} \textbf{\bibinfo{volume}{90}},
  \bibinfo{pages}{247204} (\bibinfo{year}{2003}).

\bibitem[{\citenamefont{{E. Miranda} and {J. C.
  Xavier}}(2004)}]{mirandaxavier04}
\bibinfo{author}{\bibnamefont{{E. Miranda}}} \bibnamefont{and}
  \bibinfo{author}{\bibnamefont{{J. C. Xavier}}}, \bibinfo{journal}{Physica C}
  \textbf{\bibinfo{volume}{408-410}}, \bibinfo{pages}{179}
  (\bibinfo{year}{2004}).

\bibitem[{\citenamefont{Shibata and Hotta}(2005)}]{shibatahotta05}
\bibinfo{author}{\bibfnamefont{N.}~\bibnamefont{Shibata}} \bibnamefont{and}
  \bibinfo{author}{\bibfnamefont{C.}~\bibnamefont{Hotta}},
  \bibinfo{journal}{arXiv:cond-mat/0503476v1}  (\bibinfo{year}{2005}).

\bibitem[{\citenamefont{{C. Hotta} and {N. Shibata}}(2006)}]{shibatadimer}
\bibinfo{author}{\bibnamefont{{C. Hotta}}} \bibnamefont{and}
  \bibinfo{author}{\bibnamefont{{N. Shibata}}}, \bibinfo{journal}{Physica B}
  \textbf{\bibinfo{volume}{378-380}}, \bibinfo{pages}{1039}
  (\bibinfo{year}{2006}).

\bibitem[{\citenamefont{Auerbach}(1994)}]{Auerbach}
\bibinfo{author}{\bibfnamefont{A.}~\bibnamefont{Auerbach}},
  \emph{\bibinfo{title}{Interacting {E}lectrons and {Q}uantum {M}agnetism}},
  Graduate Texts in Contemporary Physics (\bibinfo{publisher}{Spring-Verlag,
  Berlin}, \bibinfo{year}{1994}).

\bibitem[{\citenamefont{{S. R. White}}(1992)}]{white}
\bibinfo{author}{\bibnamefont{{S. R. White}}}, \bibinfo{journal}{Phys. Rev.
  Lett.} \textbf{\bibinfo{volume}{69}}, \bibinfo{pages}{2863}
  (\bibinfo{year}{1992}).

\bibitem[{\citenamefont{{S. R. White}}(1993)}]{white2}
\bibinfo{author}{\bibnamefont{{S. R. White}}}, \bibinfo{journal}{Phys. Rev. B}
  \textbf{\bibinfo{volume}{48}}, \bibinfo{pages}{10345} (\bibinfo{year}{1993}).

\bibitem[{\citenamefont{{F. D. M. Haldane}}(1982)}]{haldanedimer}
\bibinfo{author}{\bibnamefont{{F. D. M. Haldane}}}, \bibinfo{journal}{Phys.
  Rev. B} \textbf{\bibinfo{volume}{25}}, \bibinfo{pages}{4925}
  (\bibinfo{year}{1982}).

\bibitem[{\citenamefont{{K. Okamoto} and {K. Nomura}}(1992)}]{okamotonomura92}
\bibinfo{author}{\bibnamefont{{K. Okamoto}}} \bibnamefont{and}
  \bibinfo{author}{\bibnamefont{{K. Nomura}}}, \bibinfo{journal}{Phys. Lett. A}
  \textbf{\bibinfo{volume}{169}}, \bibinfo{pages}{433} (\bibinfo{year}{1992}).

\bibitem[{\citenamefont{{R. Bursill} et~al.}(1995)\citenamefont{{R. Bursill},
  {G. A. Gehring}, {D. J. J. Farnell}, {J. B. Parkinson}, {T. Xiang}, and {C.
  Zeng}}}]{bursilletal}
\bibinfo{author}{\bibnamefont{{R. Bursill}}}, \bibinfo{author}{\bibnamefont{{G.
  A. Gehring}}}, \bibinfo{author}{\bibnamefont{{D. J. J. Farnell}}},
  \bibinfo{author}{\bibnamefont{{J. B. Parkinson}}},
  \bibinfo{author}{\bibnamefont{{T. Xiang}}}, \bibnamefont{and}
  \bibinfo{author}{\bibnamefont{{C. Zeng}}}, \bibinfo{journal}{J. Phys.:
  Condens. Matter} \textbf{\bibinfo{volume}{7}}, \bibinfo{pages}{8605}
  (\bibinfo{year}{1995}).

\bibitem[{\citenamefont{{S. R. White} and {I. Affleck}}(1996)}]{whiteaffleck96}
\bibinfo{author}{\bibnamefont{{S. R. White}}} \bibnamefont{and}
  \bibinfo{author}{\bibnamefont{{I. Affleck}}}, \bibinfo{journal}{Phys. Rev. B}
  \textbf{\bibinfo{volume}{54}}, \bibinfo{pages}{9862} (\bibinfo{year}{1996}).

\bibitem[{\citenamefont{{C. K. Majumdar} and {D. K.
  Ghosh}}(1969{\natexlab{a}})}]{majumdarghosh1}
\bibinfo{author}{\bibnamefont{{C. K. Majumdar}}} \bibnamefont{and}
  \bibinfo{author}{\bibnamefont{{D. K. Ghosh}}}, \bibinfo{journal}{J. Math.
  Phys.} \textbf{\bibinfo{volume}{10}}, \bibinfo{pages}{1388}
  (\bibinfo{year}{1969}{\natexlab{a}}).

\bibitem[{\citenamefont{{C. K. Majumdar} and {D. K.
  Ghosh}}(1969{\natexlab{b}})}]{majumdarghosh2}
\bibinfo{author}{\bibnamefont{{C. K. Majumdar}}} \bibnamefont{and}
  \bibinfo{author}{\bibnamefont{{D. K. Ghosh}}}, \bibinfo{journal}{J. Math.
  Phys.} \textbf{\bibinfo{volume}{10}}, \bibinfo{pages}{1399}
  (\bibinfo{year}{1969}{\natexlab{b}}).

\bibitem[{\citenamefont{{C. K. Majumdar}}(1970)}]{majumdarghosh3}
\bibinfo{author}{\bibnamefont{{C. K. Majumdar}}}, \bibinfo{journal}{J. Phys. C}
  \textbf{\bibinfo{volume}{3}}, \bibinfo{pages}{911} (\bibinfo{year}{1970}).

\bibitem[{\citenamefont{{B. S. Shastry} and {B.
  Sutherland}}(1981)}]{Shastrysoliton}
\bibinfo{author}{\bibnamefont{{B. S. Shastry}}} \bibnamefont{and}
  \bibinfo{author}{\bibnamefont{{B. Sutherland}}}, \bibinfo{journal}{Phys. Rev.
  Lett.} \textbf{\bibinfo{volume}{47}}, \bibinfo{pages}{964}
  (\bibinfo{year}{1981}).

\bibitem[{\citenamefont{{E. Sorensen} et~al.}(1998)\citenamefont{{E. Sorensen},
  {I. Affleck}, {D. Augier}, and {D. Poilblanc}}}]{afflecksoliton}
\bibinfo{author}{\bibnamefont{{E. Sorensen}}},
  \bibinfo{author}{\bibnamefont{{I. Affleck}}},
  \bibinfo{author}{\bibnamefont{{D. Augier}}}, \bibnamefont{and}
  \bibinfo{author}{\bibnamefont{{D. Poilblanc}}}, \bibinfo{journal}{Phys. Rev.
  B} \textbf{\bibinfo{volume}{58}}, \bibinfo{pages}{R14701}
  (\bibinfo{year}{1998}).

\end{thebibliography}

\end{document}